\documentclass[twocolumn,pre,a4paper,showpacs]{revtex4}
\usepackage{amsbsy}
\usepackage{amssymb}
\usepackage[dvips]{graphicx}

\begin{document}

\title{Sparse essential interactions in model networks of gene regulation}

\author{Z. Burda}
\affiliation{Marian Smoluchowski Institute of Physics
and Mark Kac Complex Systems Research Centre, Jagellonian University,
Reymonta 4, 30-059 Krakow, Poland}

\author{A. Krzywicki}
\affiliation{Univ Paris-Sud, LPT ; CNRS,
UMR8627, Orsay, F-91405, France.}

\author{O. C. Martin}
\affiliation{Univ Paris-Sud, LPTMS ; CNRS,
UMR8626, Orsay, F-91405, France.}

\author{M. Zagorski}
\affiliation{Marian Smoluchowski Institute of Physics
and Mark Kac Complex Systems Research Centre, Jagellonian University,
Reymonta 4, 30-059 Krakow, Poland}

\date{\today}
\begin{abstract}
Gene regulatory networks typically have low in-degrees, whereby 
any given gene is regulated by few of the genes in the network.
What mechanisms
might be responsible for these low in-degrees?
Starting with an accepted framework of the 
binding of transcription factors to DNA, we consider
a simple model of gene regulatory dynamics. In
this model, we show that the constraint of having a given
function leads to the
emergence of minimum connectivities compatible with
function. We exhibit mathematically this behavior within a 
limit of our model and show that it also arises in the
full model. As a consequence, functionality
in these gene networks is parsimonious, \emph{i.e.}, 
is concentrated on a sparse number
of interactions as measured for instance by their essentiality. 
Our model thus provides a simple mechanism
for the emergence of sparse regulatory networks, and
leads to very heterogeneous effects of mutations.
\end{abstract}
\maketitle

\section{Introduction}
During the last decade, genomic studies have revealed
that complex organisms
typically do not have many more genes than less complex ones. 
Because of this, the paradigm for thinking about biological complexity 
has shifted from the number of genes to the
way they may work together: 
higher complexities might be associated with
a greater proportion of regulatory genes. In particular,
there are strong indications in eukaryotes and prokaryotes
that for increasing genome size the number of 
regulatory genes grows faster than linearly in the total
number of genes~\cite{Nimwegen2003,BabuLuscombe2004}. Hence it is 
appropriate to consider biological complexity in the framework of
interaction networks. This shift from components
to the associated interactions
has received increasing attention in many scientific communities,
with applications ranging from network biology to sociology.
The relevance of this conceptual framework for biology has 
been repeatedly emphasized (see, for example
the review~\cite{BarabasiOltvai2004})
and has benefited from inputs from other fields and from
statistical mechanics in particular~\cite{AlbertBarabasi2002}.
We will therefore freely use the network terminology, 
refering to nodes, their degrees, distinguishing between in and out degrees etc.

From studies that strive to unravel gene regulatory networks (GRN), 
several qualitative properties transpire:
(i) a given gene is generally influenced by a ``small'' number of
other genes (low in-degree of the network of interactions when
compared to the largest possible degree); 
(ii) some genes are very pleiotropic (the out-degree of some
nodes of the network can be high); (iii) GRN seem
to be robust to change (e.g. to environmental
fluctuations or to mutations), a feature
that is also found at many other levels of biological 
organisation~\cite{BornholdtSneppen2000,EdlundAdami2004,
WagnerBook2005,ChavesAlbert2005}.
A simple way to build robustness into a network is to have rather 
dense connections, effectively incorporating redundancy
in a local or global way. Furthermore, the number
of networks having $m$ interactions grows very quickly with $m$. Thus
when modeling GRN, the network realizations that perform a
given regulatory function are dominantly of very high degree.
However this is not the case experimentally, at least with respect
to the in-degree, and so models so far have had to 
build in limitations to the accessible
connectivities~\cite{KauffmanBook1993,AldanaCluzel2003,KauffmanPeterson2004}.
In this work we show that such shortcomings of models can be overcome
by taking into account the known mechanisms underlying genetic
interactions: gene regulation is mediated
via the molecular recognition
of DNA motifs by transcription factors, and this leads
to biophysical constraints on interaction strengths. Within this 
relatively realistic framework,
we shall see that networks are in fact driven to be parsimonious
(the essential interactions are sparse) for the in-degree
while the out-degree is unconstrained.

We begin by explaining the mechanisms incorporated into
the model, in particular the determinants of the
interactions. We follow standard 
practice~\cite{HippelBerg1986,BergHippel1987,GerlandMoroz2002} when
modeling interactions between DNA binding sites
and transcription factors: the affinity is taken to 
depend on the mismatch between two character strings.
We also specify how gene expression dynamics depend
on these interactions and what 
``function'' the networks must implement to be considered viable.

Before proceeding, it is perhaps useful to point out briefly
the main similarities and differences between the approach adopted
in this work and in previous literature. Our model belongs 
to the class known as ``threshold models'', used widely in 
describing neural networks~\cite{Gardner1988} and more recently 
in GRN modeling~\cite{Wagner1996}. Within 
such a framework, one represents the GRN
by its matrix of connections, and 
mutations correspond to random modifications of this matrix.
In our approach, mutations are also random of course, but we 
mimic the underlying microscopic effects of a mutation and
this forces us to work with weighted interactions;
this more realistic way of treating mutations has rather striking 
consequences as we shall see. Note that we focus on generic 
aspects of the problem, without attempting to reproduce 
specific experimental data.

After setting the general framework, we present some of the
mathematical and numerical tools
we use to analyze the model.
Results are first given for the full model, derived using
computational methods. Then we focus on 
a limiting case of the model
for which a mathematical analysis can be pushed rather far. We demonstrate
there that the constraint of having a given function 
makes the networks be marginally
``viable'' and that the corresponding connectivity is in a sense minimal,
\emph{i.e.}, networks are as sparse as they can be
subject to maintaining their function. This same
principle applies to the full model and
remarkably, the simple limit proves to be an excellent
approximation. Along with the
spontaneous appearance of sparseness, we find that the network
interactions are quite robust to 
change~\cite{NimwegenCrutchfield1999,WagnerBook2005}: only those 
few binding sites that
are ``effectively'' used are fragile, mutations of the
other (little used) binding sites have almost no effect.
Thus robustness to mutational changes is very high for
most binding sites while the ``essential'' interactions
have much lower robustness; robustness is heterogeneously distributed
in the network.
Implications of this sparseness are developed
in the discussion; in particular, a consequence
is that redundant interactions are rapidly eliminated under evolution if no new
function arises which might change the selection pressure.

\section{Framework: model of interacting genes}
\label{sec:framework}

\subsection*{Gene expression dynamics and viability ~} 

Our framework is an abstract GRN model belonging to a family
of models that has been used many times by different
authors~\cite{Wagner1996,LiLong2004,
AzevedoLohaus2006,CilibertiMartin2007a,Leclerc2008}.
We consider $N$ genes whose products 
can have regulatory influences on the same set of genes
(retroaction) and possibly also have some
``down-stream'' consequences. However the consequences of these last effects
can be ignored for our purposes since they lead to 
no feedback on the $N$ ``core'' genes. 
Call $S_j(t)$ the expression level of gene $j$ at time $t$, in practice thought
of as the concentration of the transcription factor~\cite{PaboSauer1992} 
it produces. The dynamics of the $S_j(t)$
takes place on biochemical time
scales (typically minutes). To model that,
we keep the spirit of earlier 
work~\cite{Wagner1996,AzevedoLohaus2006,CilibertiMartin2007a,Leclerc2008},
taking the genes to be either on ($S_j=1$) or off ($S_j=0$).
Furthermore, these expression levels 
are updated synchronously at discrete time steps:
to go from time step $k$ to $k+1$, we take
\begin{equation}
\label{eq:threshold_h}
S_i^{(k+1)} = H(\sum_{j=1}^{N} W_{ij} S_j^{(k)} - h) \, .
\end{equation}
Here, $h$ is a threshold and $H$ is the Heaviside 
function, $H(x)=0$ for $x \le 0$ 
and $H(x)=1$ for $x > 0$, while $W_{ij}$ denotes
the strength of the interaction that gene $j$ has on gene $i$.
A priori, the $W_{ij}$ can be arbitrary and we have no
built-in restriction on the network's connectivity.

The ``integrate and fire'' 
functional form of Eq.~[\ref{eq:threshold_h}]
is inspired by that arising in perceptrons~\cite{Gardner1988}.
The other main
family of models that have been used for modeling gene expression
dynamics involve random boolean functions of the 
inputs~\cite{KauffmanBook1993,AlbertBarabasi2000}.
Since that framework does not provide a central role for weighted
interactions,
we have not considered here the use of this second family of models.

Eq.~[\ref{eq:threshold_h}] defines
a deterministic discrete dynamical system. After possible 
transient behavior, at large $k$ the
set of expression levels $\{ S_i^{(k)} \}_{i=1,\ldots N}$ will either
go to a (time-independent) steady state or will go into a cycle
(periodic behavior). Which case arises may depend on the initial state of
the expression levels. Following the motivation~\cite{Wagner1996}
coming from early embryo development, we consider given 
the initial expression levels, $\{ S_i^{(ini)} \}_{i=1,\ldots N}$. 
Furthermore, a network will ``perform the desired function''
if and only if, starting
with the initial expression pattern,
it will lead to the desired steady-state gene expression levels; these
must also be
given a priori and correspond to the ``target'' 
$\{ S_i^{(target)} \}_{i=1,\ldots N}$. (More complicated
choices, such as limit cycles, would also be possible.)
Hereafter we say that a network is ``viable'' if it
statisfies this functional property.
Note that in our model, all genes 
are on an equal footing; it is then easy to see that the model's
properties depend not on the details of the initial and target 
patterns, but only on the number of indices $i$ where
$S_i^{(ini)} = S_i^{(target)} = 0$ and $1$. We shall show
results when these numbers
are set to their average values if each pattern is taken at random,
but the results are not sensitive to this choice.

\subsection*{Microscopic modeling of the interactions ~} 

\begin{figure}
\includegraphics[width=8cm]{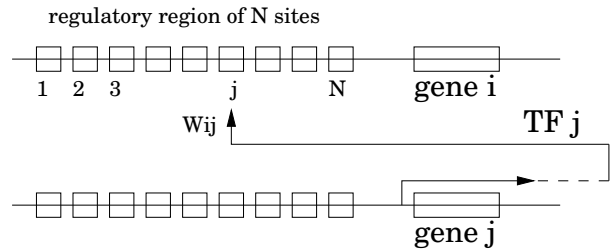}
\caption{\label{fig:model} Schematic representation of the regulatory region
  of gene $i$: there are $N$ binding sites, each labeled by an index
  $j$ ($1 \le j \le N$). Represented is the interaction $W_{ij}$ 
  mediated by the binding of TF $j$ 
  to the $j$'th site of that region. The binding affinities depend
  on the mismatch between the string of length $L$ representing the TF and that 
  representing the DNA of the corresponding binding site.}
\end{figure}

So far the framework is rather abstract, the 
interactions $W_{ij}$ are arbitrary. However much is known about
how interactions are mediated in reality, and so it
is appropriate to include this knowledge to obtain more
realistic models. To begin, the product of gene $j$
is a transcription factor, hereafter denoted 
$TF_j$, {\emph i.e.}, a protein
which modulates the rate at which other genes 
are transcribed. This modulation arises from the binding
of $TF_j$ to the regulatory region of other genes
(cf. Fig.~\ref{fig:model}).
In the absence of any bound transcription factors in its regulatory
region, gene $i$'s level of transcription will be low (considered
here as off, $S_i=0$). We allow all of the $N$ types of transcription
factors to access all the regulatory regions, but binding
depends on the affinity between the TF and the DNA content of these regions.  
To keep the model simple, we consider that each gene's
regulatory region consists of $N$ putative binding sites,
one for each of the $N$ types of TFs as illustrated
in Fig.~\ref{fig:model}.
If gene $j$ is ``on'', it will produce a certain
number $n$ of TF molecules
of type $j$; if it is off, it produces no TFs.
We shall consider different values of $n$ in our study, using
the biologically relevant range $100 \le n \le 10^4$.
(The lower value comes from the multiplicity of transcripts
in E. coli~\cite{GoldingPaulsson2005} and the expected numbers
of protein copies produced thereof, while the upper value
comes from direct measurements of numbers 
of transcription factor molecules \cite{ElfLi2007}.)

To model the affinity between TFs and binding sites, 
we follow standard
practice and represent each TF and binding site by a character 
string using a 4 letter alphabet.
The binding free-energy is then simply proportional 
to the mismatch between the two 
chains~\cite{HippelBerg1986,BergHippel1987,GerlandMoroz2002,
BergWillmann2004,MustonenKinney2008}.
This leads to the inverse Boltzmann factor
$\hat{n}_{ij} \approx C  \, e^{ \varepsilon  d_{ij}}$,
for which Gerland et al.~\cite{GerlandMoroz2002} have shown
that the constant $C$ is close to 1 and thus will be dropped
hereafter. In this formula, 
$d_{ij}$ is the Hamming distance (number 
of mismatches) between $TF_j$ and the
$j$th binding site of gene $i$; furthermore, $\varepsilon$ is the penalty 
for each mismatch (contribution to binding energy in units of $k_B T$).
Experimentally, $\varepsilon$ is inferred to have a value between one
and three if we think of each base pair of the DNA
as being represented by one
character~\cite{SaraiTakeda1989,StormoFields1998,BulykJohnson2002}. 
The number $L$ of characters
used to represent a TF or its binding site
is set using the typical number of base pairs in experimentally studied
binding sites, $10 \le L \le 15$.

When $S_j=1$, there are $n$ TF molecules of type $j$
that can bind to the $j$th site of gene $i$'s regulatory region; given that
this site can be occupied
only by one $TF$ at a time, it is common practice to take
this occupation probability to be~\cite{GerlandMoroz2002}:
\begin{equation}
\label{eq:probOccupy}
p_{ij} = \frac{1}{1 + \hat{n}_{ij} / n}  \, .
\end{equation}
For our purposes, we simplify this relation by working in the regime of
low competition as follows. First we define our $W_{ij}$ to
be proportional to the Boltzmann factor:
\begin{equation}
\label{eq:defWij}
W_{ij} = e^{-\varepsilon d_{ij}}  \; \; i,j=1,\dots,N \, .
\end{equation}
When $n$ times $\sum_{j=1}^{j=N} W_{ij} S_j$ is large enough, following
Eq.~[\ref{eq:probOccupy}], there is a 
high probability
that at least one of the binding sites in gene $i$'s
regulatory region will be occupied\footnote{
For simplicity, the model's cooperative effects arise only through the
integrate and fire equations and we do not
model repressor interactions.}.
We thus set the
threshold in Eq.~[\ref{eq:threshold_h}]
at a value $h$ which is inversely proportional to 
the number $n$ of TF molecules, $h=1/n$.
This parameter $h$ plays
a central role in the model so we shall investigate how
its value influences the behavior of the network.

\section{Methods for model analysis}
\label{sec:methods}

\subsection*{Uniform sampling of viable genotypes } 

One can think of a GRN's genes and DNA binding sites
as specifying its ``genotype''; equivalently, the genotype
can be thought of as being given by the
list of weights $W_{ij}$, corresponding to a weighted oriented graph.
Because TF are typically pleiotropic, 
they are generally thought to evolve slowly, while DNA regulatory 
regions typically have a high level of polymorphism and may evolve more 
quickly. Thus in all our study we shall consider that
the genes (and thus the TF they code)
are fixed whereas the strings of characters representing
the DNA binding sites are unconstrained. This defines the
scope of the genotype space of our model.

Now within this genotype space lies a small subset of viable genotypes,
\emph{i.e.}, thoses genotypes which lead to the correct target expression
levels given $\{ S_i^{(ini)} \}_{i=1,\ldots N}$ as starting point.
These viable GRN have the desired function or ``phenotype'';
note that the mapping from genotype to phenotype
is generally many to one. 

It is relatively straightforward to sample uniformly all
genotypes since they are specified by character strings. 
However, only a tiny part of this space corresponds to viable genotypes,
the kind we are concerned with. To sample this much smaller space,
we rely on \emph{Monte Carlo Markov Chains} (MCMC).
Within such a procedure, we start by producing (if necessary by design)
a viable GRN; then we perform a random walk in the viable subspace
of genotypes:
at each step we propose a small change of the 
characters in one of the binding sites; if the new genotype is viable, 
we accept it, otherwise we reject it and stay with the current genotype.
From this procedure, we sample uniformly
the space of viable genotypes, which allows us to
generate many random viable networks; from these unbiased samplings
we examine the statistical properties imposed by viability.
Properties include 
network sparseness, robustness to mutations and essentiality of interactions.
Furthermore, the results will depend on 
the ``specificity''~\cite{BergHippel1987,StormoFields1998,
SenguptaDjordjevic2002}
of the interactions between genes (through the alphabet size and the length
of the character strings used in our matching
process); this aspect
plays an important role in understanding how 
such networks can both function and evolve, so we will consider
how our results depend on it.

\subsection*{Model implementation}

The DNA sequences of one of our networks of $N$ genes can be represented via
$N^2$ strings of $L$ characters in a four letter
alphabet; 
indeed, a binding site for a given TF is represented by $L$ characters,
there are $N$ binding sites per gene and a total of
$N$ genes. In our framework, each TF is considered as given while
the space of all considered networks arises from 
letting the DNA sequences be variable.
Thus, instead of having an
explicit representation of the strings associated
with TFs and binding sites, it is enough to track
a binary string of length $L$ for each binding site, 
where for each entry the bit 0 (respectively 1) stands
for a mismatch (respectively a match) with the corresponding TF.
The genotype then reduces to $N^2$ such binary strings.
It is important to remember that there are 3 underlying
possible characters for each bit at 0 and only one if the
bit is 1. When using this representation for our MCMC, 
the transition rate from a 0 to a 1 bit must be three
times smaller than the transition rate from a 1 to a 0 bit.

To ensure that our MCMC is ergodic on the
time scales accessible to our computational ressources,
we use a swap operator whereby we exchange
the content of two randomly chosen neighbor binding sites.
We call ``step operation'' the following:
(1) propose successively $L$ random point mutations; (2) propose a single
swap. A ``sweep'' is then the application of $N^2$
random step operations.
The autocorrelation time of this MCMC was estimated to be less
than the time of 1 sweep.

Concerning the choices for $S^{(ini)}_{i=1,\dots,N}$ and 
$S^{(target)}_{i=1,\dots,N}$: 
if drawn at random, the fraction of terms set to 1 would be approximately 
equal to that set to 0 when $N$ is large. To 
reduce finite size effects, we force the equality
at our values of $N$ which are 
multiples of 4. Because of the permutation 
symmetry of the model, one can
always permute the indices so that $S^{(ini)}_i = 1$ 
for $i \leq N/2$ and 0 otherwise;
furthermore we also impose without loss of generality
$ S^{(target)}_i=1$ for $N/4 < i \leq 3N/4$ and 0 otherwise.
Notice that $\sum_i S^{(ini)}_i = \sum_i S^{(target)}_i = N/2$
and $\sum_i S^{(ini)}_i S^{(target)}_i = N/4$.

Finally, we need to start the MCMC with a viable GRN. 
To generate an initial genotype, we first set
\begin{equation}
W_{ij} = S^{(target)}_i S^{(target)}_j  \; \; i,j=1,\dots,N
\end{equation}
and then construct the bit strings of the binding sites by taking values
that approximate this equation. In practice this initial setting nearly
always
leads to a viable genotype; if not, other approximations are tried. From this
procedure an initial viable genotype is constructed and then the MCMC can
begin.
\par

\section{Results}
\label{sec:results}

\subsection*{Some qualitative properties ~}
\label{subsec:qualitative}

The total number of genotypes is $4^{L N^2}$; since
realistic values of $L$ are at least $10$, this number
is astronomical even for rather modest $N$. However only a tiny
fraction of these genotypes are viable. Naively, since we want
the gene expression pattern in the steady state to be 
given by $S^{(target)}$, and since there are $2^N$ possible
patterns, one may expect only a fraction of order $2^{-N}$ of
the genotypes to be viable. In fact,
the fraction is even smaller, especially as
$h$ grows. Thus if one seeks to generate random
viable GRN by producing random genotypes (strings of characters
for the binding sites), most attempts
will be unsuccessful and it will be near impossible to 
sample the space of 
viable GRNs when $N$ is $3$ or more. This is why we
relied on Monte Carlo Markov Chains to perform the sampling
of viable networks. (Such an approach is computationally efficient
if the Markov Chain has a short auto-correlation time,
which is the case here as mentioned in the methods section.)
Although it is difficult to derive properties of viable genotypes,
\emph{general} genotypes are relatively easy to understand
because there is no viability constraint. The
statistical properties of the interaction strengths $W_{ij}$ 
follows from the distribution of the mismatch between 
the character strings for a binding site and a TF. Each of the 
$L$ characters of a binding site gives a mismatch with probability
$3/4$; the total mismatch $d$ thus follows the binomial
distribution of mean $3L/4$:
\begin{equation}
\label{eq:binom}
p(d) = {L\choose d} (1/4)^{L-d} (3/4)^{d}
\end{equation}

\subsection*{Sparse essential interactions}
\label{subsec:sparse}

In contrast to general genotypes, viable genotypes 
are subject to the constraint of reaching (after
some transients) the steady
state expression given by  $\{ S_i^{(target)} \}_{i=1,\ldots N}$
when initialized in the state $\{ S_i^{(ini)} \}_{i=1,\ldots N}$.
How do the set of interactions $W_{ij}$ differ when comparing
viable and general networks?
We first address this question computationally by considering 
statistical properties of random genotypes, subject to
being viable or not.

\begin{figure}[t]
\includegraphics[width=8cm]{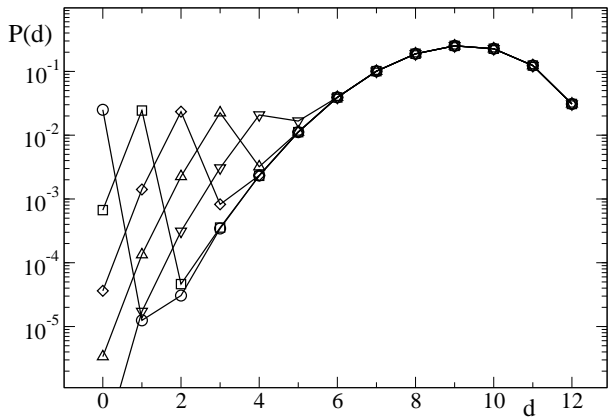}
\caption{\label{fig:mismatch} Distribution of the Hamming distance 
between a TF and the receiving DNA site for $N=20, L=12, \varepsilon=1.5$ and 
for various values of the threshold parameter: (circle) $h=0.3$, (square)
$h=0.1$, (diamond) $h=0.02$, (triangle up) 
$h=0.005$, (triangle down) $h=0.001$. The lines are to guide the eye.}
\end{figure}

The main control parameter is the threshold value $h$,
which must be compared to the typical value of 
$\sum_j W_{ij} S_j$ in the absence of the viability constraint;
denote this value by $\omega$. Representing the averages over
all genotypes by $\langle \, \rangle$, we have
$\omega = N \langle W_{ij} \rangle /2$. For this formula we have used
the fact that because of the symmetry of the 
model in the absence of the viability constraint,
the $\langle W_{ij} \rangle$ are independent of $i$ and $j$;
we have also used 
$\langle S_i \rangle = 1/2$ for random initial
and target expression states. Now when $h$ is significantly
larger than $\omega$, nearly all
random genotypes will have their expression levels
go to 0 at large times and thus will
not be viable. If a genotype is viable, it must be that 
$\sum_j W_{ij} S_j^{(target)}$ is anomalously large for
all $i$ such that $S_i^{(target)}=1$.
When this happens for such a row $i$, one may have one
entry $j^*$ for which $W_{ij^*}$ is very large, or one may spread
out the ``burden'' among several interactions
$W_{ij_1}, W_{ij_2}, \ldots W_{ij_k}$ where each weight
is a bit larger than average. In the space of all viable genotypes,
does one more often resort to the first or second strategy?
To find out, we begin by considering the distribution of the 
mismatch between TFs and their binding sites in viable networks.
We denote the mismatch by $d$; it is perhaps useful to regard
the quantity $L-d_{ij}$ as a measure of the affinity
between a TF and the corresponding DNA binding site in this
model. In Fig.~\ref{fig:mismatch} we show the distribution of
$d$ when there are $N=20$ genes, for increasing values of
the threshold parameter $h$. At the low values of $h$,
$d$ has a binomial distribution with a peak
near $d=3L/4$ as expected. However one observes at small $d$
significant deviations. In fact, as $h$ increases, 
the viability constraint become marked and 
the distribution becomes bimodal: a peak
appears at low mismatch values. Note that 
this peak shifts as $h$ increases, indicating
that there are nearly perfect matches that appear
in that regime.

\begin{figure}[t]
\includegraphics[width=8cm]{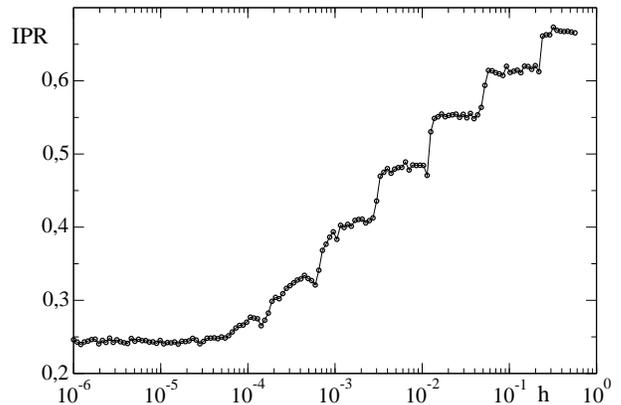}
\caption{\label{fig:IPR}  The mean of the inverse participation ratio per line
(cf. Eq.~[\ref{ipr}]) versus $h$ for $N=20, L=12, \varepsilon=1.5$.
The line is to guide the eye.  }
\end{figure}

To distinguish the two scenarios, namely whether the burden in each row $i$ is
concentrated on one vs. on multiple $j$ indices, it is 
natural to consider
the inverse participation ratio (IPR), a commonly used measure
of how many terms effectively participate in a weighted list.
The straightforward use of this approach for the 
matrix $\{W_{ij}\}$ is the IPR
$I = \sum_{i,j} W_{ij}^2 / (\sum_{i,j} W_{i,j} )^2$.
If many elements effectively participate, the value will 
be $O(1/N^2)$, while if only a few contribute per line the value will be 
not much less than $1/N$. However this index is a poor indicator
of sparseness for several reasons. First, in
the absence of the viability constraint and when $\varepsilon$ is
of order unity, the value is found to be of 
$O(1)$! The reason can be traced to the distribution of the 
$W_{ij}$ when the mismatches are binomially distributed:
because the weights $W_{ij}$ are exponential in the mismatches,
$I$ is often dominated by one or two interactions. Second,
we are actually more interested in what happens for each row
$i$ such that $S_i^{(target)}=1$; the other rows
don't need good matches and just add noise to $I$.
Thus it is appropriate to focus on the IPR restricted to 
one such line at a time. We therefore define
\begin{equation}
A_{ij} = S^{(target)}_i S^{(target)}_j (d_{ij} - 3L/4)^2 
\label{matrixA}
\end{equation}
and the associated IPR for the $i$-th line
\begin{equation}
I_i = \frac{\sum_{j}  A_{ij}^2}
{(\sum_{j} A_{ij})^2}  \; .
\label{ipr}
\end{equation}
Then we average these $I_i$'s over $i$'s for which $S_i^{(target)}=1$.
This is the IPR quantity 
plotted in Fig.~\ref{fig:IPR}. Except at
very low values of $h$, only one or two 
weights per row are significantly larger than the others in that same row.
Note that the ``staircase'' structure of the plot
is a consequence of the discrete nature of the mismatch value.

\begin{figure}
\includegraphics[width=8cm, angle=0]{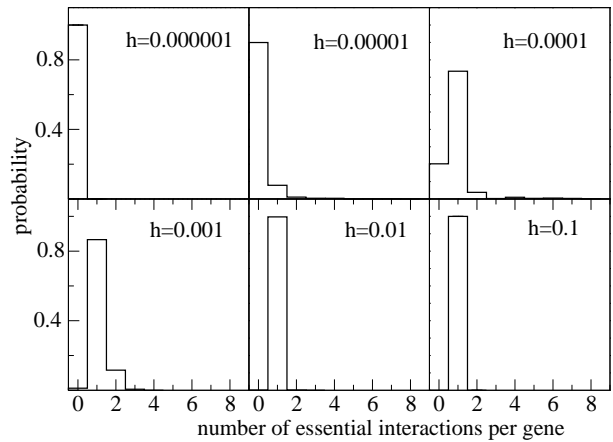}
\caption{\label{fig:essential} Probability distribution 
of the number of essential interactions
per row of the matrix specifying a viable network for $N=20, L=12, 
\varepsilon=1.5$ and a range of 
values of $h$.}
\end{figure}

It is possible to further explore the statistical properties
of viable genotypes by considering not
the weights themselves but their function.
One can ask whether ``essential'' interactions are sparse
(few essential interactions per row), \emph{i.e.},
in a given viable genotype, how many of its interactions
have the property that viability is lost when the interaction is removed
($W_{ij}$ is set to $0$).
We find that as soon as $h$ is not too small, 
there is almost always just one essential
interaction per gene as shown in 
Fig.~\ref{fig:essential} for $N=20$ and $L=12$. The same 
result holds for other
relevant values of $N$ and $L$, suggesting that
within our models,
the drive towards sparse interactions arises in 
regimes of biological relevance.
We also considered a stronger measure
of essentiality: we asked that the viability be lost when 
the interaction's mismatch
is increased by one. Remarkably, the rule ``one essential
interaction per gene'' generally held here too. Thus mutations
in these interactions are typically deleterious, while 
mutations in the vast majority of the other interactions have
no consequence on viability. This shows that mutational
robustness is very heterogeneously distributed among the 
interactions in the network.

\subsection*{Mathematical analysis in a simple limit ~}
\label{subsec:mathematical}

As already mentioned, under the dynamical process Eq.~[\ref{eq:threshold_h}],
when the expression levels reach the 
target, the dynamics must be at a fixed point if the 
genotype is viable. Then the
different lines of the matrix $\{W_{ij}\}$ have to satisfy the
``fixed point'' constraint
and in each line it is sufficient to consider those elements which are
multiplied by 1. It is therefore worth considering a toy model 
where transient effects in the dynamics are neglected. In such a framework, 
we consider only the fixed point conditions
that now can be thought of as coming from the particular choice
$S_i^{(ini)} = S_i^{(target)}$ for each $i$. For each such
index $i$ (or equivalently line of the matrix $\{W_{ij}\}$),
the toy model leads to the partition function
\begin{equation}
Z_{toy}^{(K)}(h) = \sum_{d_1,\dots, d_K} p(d_1, d_2, \dots, d_K) 
H(\sum_{j=1}^K e^{-d_j \varepsilon} - h)
\label{ztoy}
\end{equation}
where $K=N/2$ and $H$ is the Heaviside function. 
We have also assumed that $i$ is such that
$S_i^{(target)} = 1$. For the other lines (for which
$S_i^{(target)} = 0$), if $h$  is not too small the fixed point condition
will nearly always be satisfied and so can be ignored. Because in this toy
model the different lines are independent, we can focus on
one line at a time, in line with what arises when analyzing
fixed points in neural network systems~\cite{Gardner1988}.

In this reduced problem, the state space is a 
K-dimensional hypercube ${\cal C}$ with edge length $L$, 
${\cal C} : (d_1, d_2, \dots, d_K), \; d_j = 0, 1, \dots, L$
being the $j$th mismatch. The a priori mismatch probability in fact
factorizes:
\begin{equation}
p(d_1, d_2, \dots, d_K) = \prod_{i=1}^K p(d_i)
\label{prob}
\end{equation}
with $p(d)$ given by Eq.~[\ref{eq:binom}]. From this we see that
$[Z_{toy}^{(K)}(h)]^K$ gives the fraction of random genotypes
that are viable.
Notice, that for $L$ large and for $d$ small enough 
\begin{equation}
p(d) \sim \frac{L^d}{4^L} \ll 1 , \; \; d \ll L \; .
\label{smallness}
\end{equation}

We wish to understand the effect of the Heaviside constraint on 
the probability distribution of the mismatches and
compare with what happens in the full model. We first
define $d_h \equiv d_h(\varepsilon, h)$ by the equation 
$e^{-\varepsilon d_h}= h$ and assume that $h$ is such that
Eq.~[\ref{smallness}] holds for $d<d_h$. Now remove from the state space 
${\cal C}$ the sub-hypercube ${\cal C'}$: $d_i \geq d_h$ for 
all $i=1,\dots , K$. In 
this reduced state space we keep the same probabilities up to a 
normalization factor:
\begin{eqnarray}
\lefteqn{\tilde{p}(d_1, d_2, \dots, d_K) =} \nonumber \\
&& p(d_1) \dots p(d_K)
 \left[ \prod_j \big(1 - H(d_h - d_j)\big) \right] /A \; .
\label{newprob2}
\end{eqnarray}
The factor in brackets is 1 within the reduced
state space and vanishes outside of it, so it serves simply to
filter out elements in ${\cal C'}$. Also,
\begin{eqnarray}
A & = & 1 - \sum_{d_1,\dots,d_L} \prod_{j=1}^K p(d_j) H(d_j - d_h) \nonumber \\
  & = & 1 - \big(1 - \sum_m p(m) H(d_h-m)\big)^K
\label{norm}
\end{eqnarray}
Eqs.[\ref{newprob2}-\ref{norm}] are exact. To proceed further 
we take advantage of Eq.~[\ref{smallness}]: in the r.h.s. of 
Eqs.~[\ref{newprob2}-\ref{norm}] we expand when possible the products and 
drop all quantities where $p(d)$'s with $d<d_h$ appear at higher order 
than first, \emph{e.g.}, $\dots p(d_i)p(d_j)\dots$ with $d_{i,j}<d_h$. This yields 
\begin{eqnarray}
\lefteqn{\tilde{p}(d_1, d_2, \dots, d_K) \approx }\nonumber \\
&& p(d_1) \dots p(d_K) 
 \sum_{j=1}^K H(d_h - d_j)/K \sum_{m<d_h} p(m) 
\label{newprob3}
\end{eqnarray}
The marginal distribution of a mismatch, say $d_1$, is obtained by summing over all $d_i$ 
with $i>1$. (Note that the marginals distributions do not depend on
the value of the index.) Changing notation $d_1 \to d$ one easily obtains
\begin{equation}
\tilde{p}(d) \approx
\frac{K-1}{K} p(d) + \frac{1}{K} \frac{p(d) H(d_h - d)}{\sum_{m<d_h}
  p(m)} \; .
\label{eq:marginal}
\end{equation}
Eq.~[\ref{eq:marginal}] has a very simple interpretation: 
in the reduced state space, 
the shape of the probability distribution of any mismatch 
is essentially that without the viability constraint, but with
an additional peak at small values of the mismatch. 
There is thus one ``leading'' mismatch taking 
care of most of the constraint, while 
the other mismatches behave approximately as if they were unconstrained. 
Such a behavior is exactly
what we saw happened in the full model. In brief, the
effect of the viability constraint condenses on one of the 
entries of the row considered, the other entries behave as
if there were no constraint. Furthermore, one has sparseness
of the essential interactions and a high IPR. This situation,
where the IPR goes from low to high values as a parameter ($h$ here)
is increased, is reminiscent of many ``condensation'' phase transitions. 
In a biological context, such a transition has been observed in
another genetic system, but based on 
epistatic interactions~\cite{NeherShraiman2009}
and otherwise unrelated to our framework. It has
also been seen in statistical physics problems~\cite{BialasBurda1997}.

\begin{figure}
\includegraphics[width=8cm]{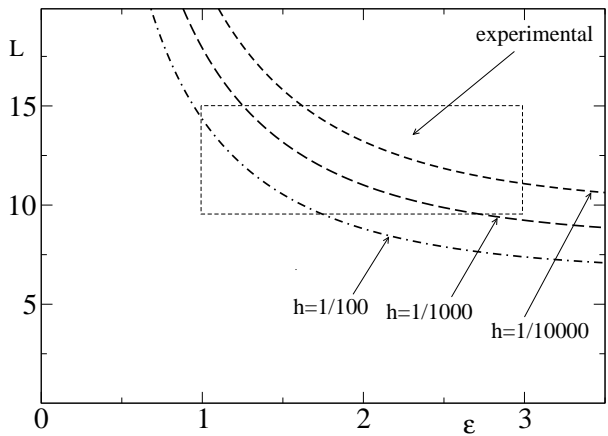}
\caption{\label{fig:phaseDiagram}  The curves show the lower 
limit of the region
where $h = 10 (N/2)[0.25+0.75 exp(-\varepsilon)]^L$.  }
\end{figure}

In Eq.~[\ref{eq:marginal}], the neglected terms are order $K p(d)$ smaller than those kept. Hence 
to justify dropping those terms we must have
\begin{equation}
KL^{d_h}/4^L \sim KL^{[\ln(1/h)/\varepsilon]}/4^L \ll 1 \; .
\label{error}
\end{equation}
The constraint imposed so far on the state space 
(exclusion of ${\cal C'}$) leads to simple formulae but it is stronger than 
the one imposed by the toy partition function, namely
\begin{equation}
\sum_{j=1}^K e^{-\varepsilon d_j} > h \; .
\label{constraint}
\end{equation}
If we replace in this relation the inequality by an
equality and assume that the $d_j$'s are continuous variables, we obtain the
definition of a hypersurface, call 
it ${\cal S}$, which is included in ${\cal C'}$.
From this we see that the constraint Eq.~[\ref{constraint}]
removes not the whole hypercube ${\cal C'}$ 
but only the points lying beyond ${\cal S}$.
The perturbative relation Eq.~[\ref{eq:marginal}] is nevertheless an
excellent approximation provided the probabilities associated 
with those points of ${\cal C'}$
which remain in the state
space are very small. This is usually the
case for values of parameters we consider and that are of
biological interest. In fact, the quality of the approximation
could have been guessed given the form of the 
data shown in Fig.\ref{fig:mismatch}.


\section{Discussion and conclusions}
\label{sec:discussion}

We considered a fairly general model of a gene regulatory network (GRN)
in which function is identified with reaching given 
target gene expression levels.
By simulation and mathematical analysis, we investigated the properties
of networks in this model under the constraint that they be ``viable''
(\emph{i.e.}, have the desired function). We find 
that for a certain range of values 
of the model's parameters,
the viability constraint leads to sparse GRN;
we have quantified this through both inverse participation
ratios of the interactions and through the 
sparsity of ``essential'' interactions. Interestingly,
the effects of the viability constraint condense onto just a few of the 
possible binding sites, the others being non-functional.
As a result, nearly all mutations of the binding sites have no effect
on the viability and so such sites have a very high mutational 
robustness. However, for those few sites which bear the 
burden of the constraint, 
the majority of mutations are deleterious so their mutational robustness
is low. Thus in our GRN, the mutational robustness is extremely
heterogeneous from site to site. In addition, any ``redundant''
interaction is expected to become lost under evolutionary dynamics since
mutations will remove it and condense the burden of viability
onto a smaller number of interactions.

Although our modeling of the regulation of gene expression is relatively 
idealized (cf. the simple dynamical process Eq.~[\ref{eq:threshold_h}]),
other features of the model presented in this paper are 
fairly realistic; in particular we have insisted on including
interactions through the biophysical mechanism of molecular recognition
and affinity. It is therefore interesting 
that the sparseness of GRNs comes out very 
naturally in this framework. This is well 
illustrated by our numerics and by the analytic 
calculation in the previous section. It should be clear that 
this sparseness is a result of the combined 
effect of several causes: the viability constraint, the low probability 
of a small mismatch between TF and the binding site 
of DNA, the size $L$ of this segment, 
the not-too-small spacing (in units of $k_B T$) between the energy 
levels that determine the strength of TF-DNA interactions, 
and finally the value of the threshold $h$ itself in the 
expression dynamics of Eq.~[\ref{eq:threshold_h}].
A necessary condition for our GRNs to be sparse is that this
threshold $h$ be significantly larger than 
the total strength contributed by random 
gene interactions in the absence of the viability constraint. For
a four letter code, assuming, as we do, that 
the binding energy is additive in mismatches 
and that every mismatch costs the same, one 
gets 
the condition
\begin{equation}
h \gg \frac{N}{2}(0.25+0.75 e^{-\varepsilon})^L
\label{limits}
\end{equation}
Note that within a two letter code, the condition forces one to
larger values of $L$ (close to 20) and thus 
beyond what is realistic biologically.

Since the model parameters correspond to 
measurable quanties, it is appropriate to compare to biological values.
According to Eq.~[\ref{eq:probOccupy}]
the probability that a TF occupies a DNA site
is controlled by $n$, the number of these 
molecules. Thus, this probability is greater
than $1/2$ when
the corresponding Boltzmann factor is larger 
than $1/n$. Hence, $1/n$ is an estimate of our
threshold $h$: the reasonable range is roughly 
$1/10000 < h < 1/100$. Interpreting $\gg$ as
``larger by one order of magnitude'', \emph{i.e.}, by
a factor 10, one gets an allowed region in parameter
space as illustrated in Fig.~\ref{fig:phaseDiagram}.
The predicted domain of relevance is above the
corresponding curves (taken at $N=20$ and illustrative values
of $h$). We see that $\varepsilon$ and $L$ should not be
too small. Moreover, it is gratifying 
that the experimental range of these parameters 
(indicated by a rectangle)
is near the border and, most of it, within 
this region. The model would remain meaningful if
$L$ and $\varepsilon$ were even larger. However, in the analogue 
of Fig.~\ref{fig:mismatch}, the point at $d=0$ would 
dominate strongly over the few neighboring
$d$ points, and the system would be robust but not 
evolvable~\cite{BergWillmann2004,MustonenKinney2008}.
Presumably, both to reach functional GRN and
to allow these to be evolvable, it is desirable not to 
be too dominated by essential interactions, so
the probability of having two co-dominant mismatches should not 
be completely negligible.
Such effects go beyond our model as we work within
a ``strong selection'' limit: a GRN is viable or not,
no graduation is allowed; when a continuous fitness 
replaces viability, evolvability should be strongly enhanced.

It is worth emphasizing that as
the number $N$ of genes grows, it is necessary
to increase slowly either $L$, $\varepsilon$ or $h$.
$\varepsilon$ is constrained by biophysical processes and thus not
evolvable, and $L$ seems the best candidate for the system
to adapt to increasing $N$~\cite{SenguptaDjordjevic2002}. Note 
nevertheless that the
effects of growing $N$ are mild and that in practice
regulation is modular, so effectively biological GRN
have only modest values of $N$.
Finally, as already mentioned in the introduction, our model
belongs to the class of threshold models that have a much wider
applicability than GRN. Therefore, the emergence of essential interactions
following the mechanisms outlined in this paper is expected in all cases where
the typical magnitude of the ``local field'' $\sum_i W_{ij} S_j$ 
is small compared to the threshold.

\begin{acknowledgments}
We thank V. Hakim, the late P. Slonimski and A. Wagner for helpful comments.
This work was supported
by the EEC's FP6 Marie Curie RTN under
contract MRTN-CT-2004-005616 (ENRAGE: European
Network on Random Geometry), 
by the EEC's IST project GENNETEC - 034952, by the
Marie Curie Actions Transfer of Knowledge project ``COCOS'',
Grant No. MTKD-CT-2004-517186, and by the Polish Ministry of Science
Grant No.~N~N202~229137 (2009-2012).
The LPT and LPTMS are Unit\'e de Recherche de
l'Universit\'e Paris-Sud associ\'ees au CNRS.
\end{acknowledgments}



\end{document}